\documentclass[12pt,preprint]{aastex}

\slugcomment{}

\shorttitle{Lensed QSO SDSS J0246$-$0825}
\shortauthors{Inada et al.}

\begin{document}

\title{SDSS~J0246$-$0825: A New Gravitationally Lensed Quasar 
from the Sloan Digital Sky Survey\footnotemark[1]}

\author{
Naohisa Inada,\altaffilmark{2}
Scott Burles,\altaffilmark{3}
Michael D. Gregg,\altaffilmark{4,5}
Robert H. Becker,\altaffilmark{4,5}
Paul L. Schechter,\altaffilmark{3}
Daniel J. Eisenstein,\altaffilmark{6}
Masamune Oguri,\altaffilmark{7,8}
Francisco J. Castander,\altaffilmark{9}
Patrick B. Hall,\altaffilmark{10}
David E. Johnston,\altaffilmark{8}
Bartosz Pindor,\altaffilmark{11}
Gordon T. Richards,\altaffilmark{8}
Donald P. Schneider,\altaffilmark{12}
Richard L. White,\altaffilmark{13}
J. Brinkmann,\altaffilmark{14}
Alexander S. Szalay,\altaffilmark{15}
and Donald G. York\altaffilmark{16,17}
}

\footnotetext[1]{Based on observations with the NASA/ESA {\em Hubble Space Telescope}, 
obtained at the Space Telescope Science 
Institute, which is operated by the Association of Universities for Research 
in Astronomy, Inc., under NASA contract NAS 5-26555. These observations are 
associated with {\em HST} program 9744. }

\altaffiltext{2}{Institute of Astronomy, Faculty of Science, University of Tokyo, 2-21-1 Osawa, Mitaka, Tokyo 181-0015, Japan.}
\altaffiltext{3}{Physics Department, Massachusetts Institute of Technology, 77 Massachusetts Avenue, Cambridge, MA 02139.}
\altaffiltext{4}{IGPP-LLNL, L-413, 7000 East Avenue, Livermore, CA 94550.}
\altaffiltext{5}{Physics Department, University of California, Davis, CA 95616.}
\altaffiltext{6}{Steward Observatory, University of Arizona, 933 North Cherry Avenue, Tucson, AZ 85721.}
\altaffiltext{7}{Department of Physics, University of Tokyo, Hongo 7-3-1, Bunkyo-ku, Tokyo 113-0033, Japan.}
\altaffiltext{8}{Princeton University Observatory, Peyton Hall, Princeton, NJ 08544.}
\altaffiltext{9}{Institut d'Estudis Espacials de Catalunya/CSIC, Gran Capita 2-4, 08034 Barcelona, Spain.}
\altaffiltext{10}{Department of Physics \& Astronomy, York University, 4700 Keele St., Toronto, ON M3J 1P3, Canada.}
\altaffiltext{11}{Department of Astronomy, University of Tronto, 60 St. George Street, Tronto, Ontario M5S 3H8, Canada.} 
\altaffiltext{12}{Department of Astronomy and Astrophysics, The Pennsylvania State University, 525 Davey Laboratory, University Park, PA 16802.} 
\altaffiltext{13}{Space Telescope Science Institute, 3700 San Martin Drive, Baltimore, MD 21218.}
\altaffiltext{14}{Apache Point Observatory, P.O. Box 59, Sunspot, NM 88349.}
\altaffiltext{15}{Center for Astrophysical Sciences, Department of Physics \& Astronomy, Johns Hopkins University, Baltimore, MD 21218.}
\altaffiltext{16}{Department of Astronomy and Astrophysics, The University of Chicago, 5640 South Ellis Avenue, Chicago, IL 60637.}
\altaffiltext{17}{Enrico Fermi Institute, The University of Chicago, 5640 South Ellis Avenue, Chicago, IL 60637.}

\begin{abstract}
We report the discovery of a new two-image gravitationally lensed quasar,
SDSS~J024634.11$-$082536.2 (SDSS~J0246$-$0825). This object was selected
as a lensed quasar candidate from the Sloan Digital Sky Survey (SDSS) 
by the same algorithm that was used to discover other SDSS lensed quasars 
(e.g., SDSS~J0924+0219). 
Multicolor imaging with the Magellan Consortium's Walter 
Baade 6.5-m telescope and the spectroscopic observations using the W. M. 
Keck Observatory's Keck II telescope confirm that SDSS~J0246$-$0825 
consists of two lensed images ($\Delta{\theta}=$1\farcs04) of a source quasar 
at $z=1.68$. Imaging observations with the Keck telescope and the {\sl Hubble Space Telescope} 
reveal an extended object between the two quasar components, which is likely to be a lensing
galaxy of this system. From the absorption lines in the spectra of quasar
components and the apparent magnitude of the galaxy, combined with the
expected absolute magnitude from the Faber-Jackson relation, we 
estimate the redshift of the lensing galaxy to be $z=0.724$. 
A highly distorted ring is visible in the {\sl Hubble Space Telescope}
images, which is likely to be the lensed host galaxy of the source quasar.
Simple mass modeling predicts the possibility that there is a small (faint) 
lensing object near the primary lensing galaxy. 
\end{abstract}

\keywords{gravitationally lensing --- 
quasars: individual (SDSS~J024634.11$-$082536.2)}

\section{Introduction}

Since the discovery of the first lensed quasar, 
Q0957+561 \citep*{walsh79}, 
approximately 80 lensed quasars have been discovered to date.
Gravitationally lensed quasars are now known to be an 
useful tool for cosmology and astrophysics, such as measurements of dark 
energy \citep*{turner90,fukugita90,chiba99,chae02}, 
direct probes of the Hubble constant
\citep{refsdal64,bernstein99,koopmans03b}, and the study of lensing 
galaxies 
\citep*{kochanek91,keeton01a,oguri02,koopmans03a,rusin03,ofek03}. 
In order to use lensed quasars as these cosmological and astrophysical 
tests, large and homogeneous samples of gravitational lenses are
necessary, either to ensure good statistical precisions and to
provide sufficient number of lensing systems which are
suitable for measurements of time delays and/or for studies of 
the lensing galaxies. Although several previous
large and homogeneous lensed quasar surveys, e.g., Jodrell/VLA
Astrometric Survey and the Cosmic Lens All-Sky Survey
\citep{myers03,browne03} and  
the {\sl Hubble Space Telescope} Snapshot Survey 
\citep{maoz92,maoz93a,maoz93b,bahcall92}, are of considerable value, 
discoveries of more lensed quasars in larger surveys, 
such as the Sloan Digital Sky Survey 
\citep[SDSS;][]{york00,stoughton02,abazajian03,abazajian04}
will greatly improve the accuracy of these cosmological 
and astrophysical tests. The final number of spectroscopically confirmed 
quasars in the SDSS is expected to be about $10^5$, thus, if we assume the 
lensing rate of 0.1\% \citep*{turner84}, the expected number of gravitationally 
lensed quasars in the SDSS will be approximately $10^2$; the SDSS could double
the number of lensed quasars. Indeed, several newly
discovered lensed quasars in the SDSS data have already been reported 
\citep*{inada03a,inada03b,inada03c,morgan03,johnston03,pindor04,oguri04a,oguri04b}. 

In this paper, we report the discovery of another lensed quasar, 
SDSS~J024634.11$-$082536.3 (SDSS~J0246$-$0825). This object was 
selected as a lensed quasar candidate (an ``extended'' quasar) 
from about 50,000 SDSS quasars using the algorithm described in 
\citet{inada03b}, \citet{pindor04}, \citet{oguri04a} and \citet{oguri04b}. 
Since the SDSS imaging and spectroscopic data of
SDSS~J0246$-$0825 are unresolved, we conducted additional observations using
the W. M. Keck Observatory's Keck II telescope, the Magellan Consortium's
Walter Baade 6.5-m telescope (WB6.5m)\footnote[17]{The first telescope of the Magellan Project; 
a collaboration between the Observatories of the Carnegie Institution of Washington 
(OCIW), University of Arizona, Harvard University, University of Michigan, 
and Massachusetts Institute of Technology (MIT) to construct two 6.5 Meter optical 
telescopes in the southern hemisphere.}, and the {\sl Hubble Space
Telescope} ({\sl HST}).  These observations
demonstrate that SDSS~J0246$-$0825 is a two-image lens (the image 
separation is 1\farcs04)
of a source quasar at $z=1.68$, and that the likely lensing galaxy has been detected.

The structure of this paper is as follows.
Section \ref{sec:obs} describes both the photometric and
spectroscopic follow-up observations. In \S \ref{sec:model}, we show 
mass modeling of this lensing system. 
Finally, we present a summary and give a conclusion in \S \ref{sec:conc}.
We assume a Lambda-dominate cosmology (Spergel et al. 2003; matter density 
$\Omega_M=0.27$, cosmological constant $\Omega_\Lambda=0.73$, and the Hubble constant 
$H_{0}=70$ km sec$^{-1}$ Mpc$^{-1}$), and we use AB magnitude system \citep{oke83, fukugita96}
throughout the paper. 

\section{Follow-up Observation and Data Analysis}\label{sec:obs}

\subsection{SDSS Observations}

The SDSS is a project to conduct in parallel a photometric survey 
\citep*{gunn98,lupton99} and a spectroscopic survey 
\citep{blanton03} of 10,000 square degrees of the sky 
centered approximately on the North Galactic Pole, using dedicated 
wide-field ($3^{\circ}$ field of view) 2.5-m telescope 
at Apache Point Observatory in New Mexico, USA. 
The astrometric positions are accurate to better than about 
$0\farcs1$ rms per coordinate \citep{pier03} and the photometric 
errors are less than about 0.03 magnitude \citep{hogg01,smith02,ivezic04}.
Photometric observations are conducted using five optical broad bands
\citep{fukugita96}. After automated data processing by the 
photometric pipeline \citep{lupton01}, quasar, 
galaxy and luminous red galaxy candidates are selected by the 
target selection pipelines \citep{eisenstein01,richards02,strauss02}.
Spectra of these candidates are obtained with a
multi-fiber spectrograph covering 3800{\,\AA} to 9200{\,\AA} 
(resolution R $\sim$ 1800). 

\begin{figure}
\plotone{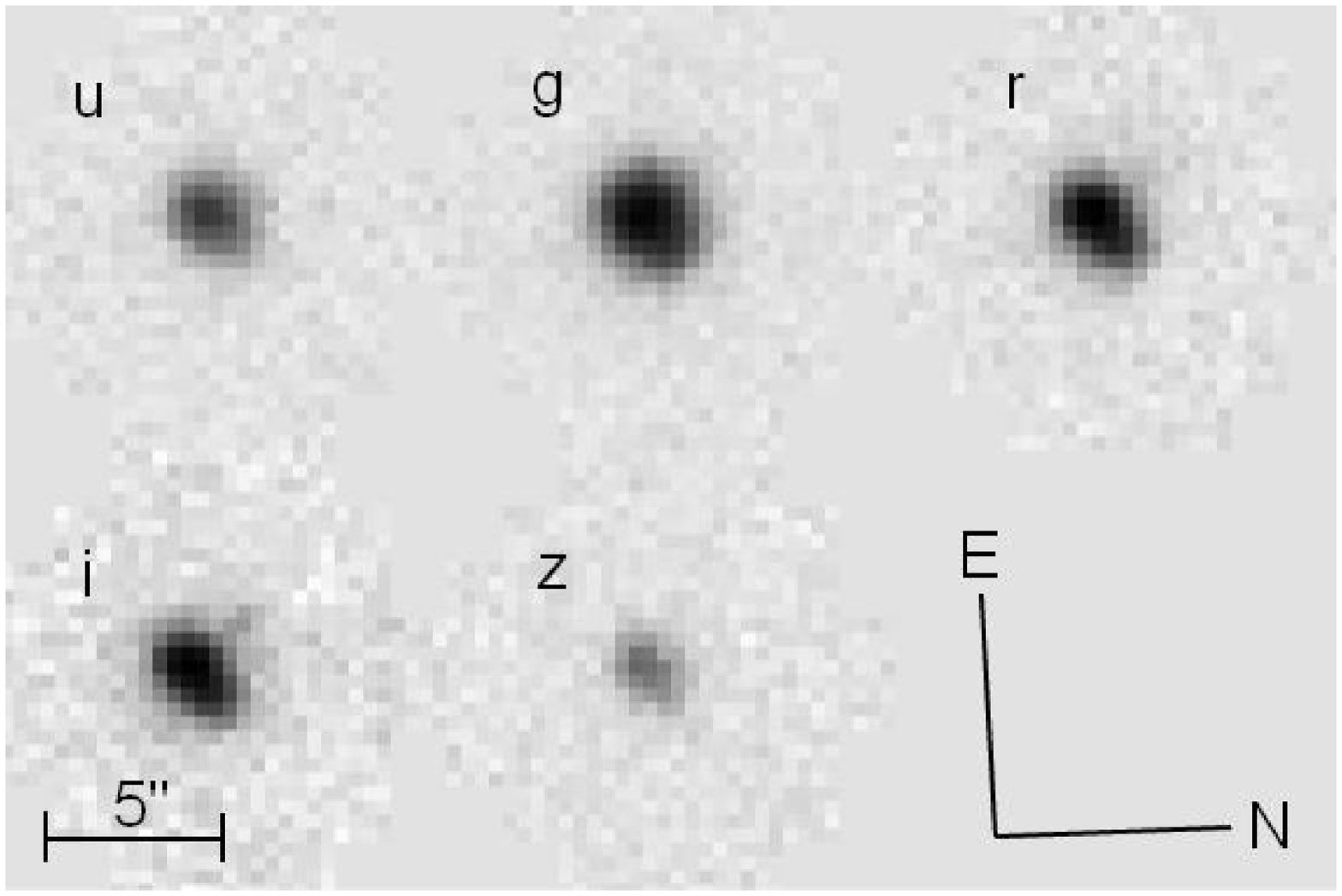}
\caption{The images of SDSS J0246-0825 in all five SDSS bands ($ugriz$). The pixel
size is $0\farcs396$ and the seeing (FWHM) of the field was about $1\farcs5$.
Total magnitudes in each band are 18.21, 17.99, 17.95, 17.60, and 17.58
in $u$, $g$, $r$, $i$, and $z$, respectively. 
\label{fig:sdss}}
\end{figure}

\begin{figure}
\plotone{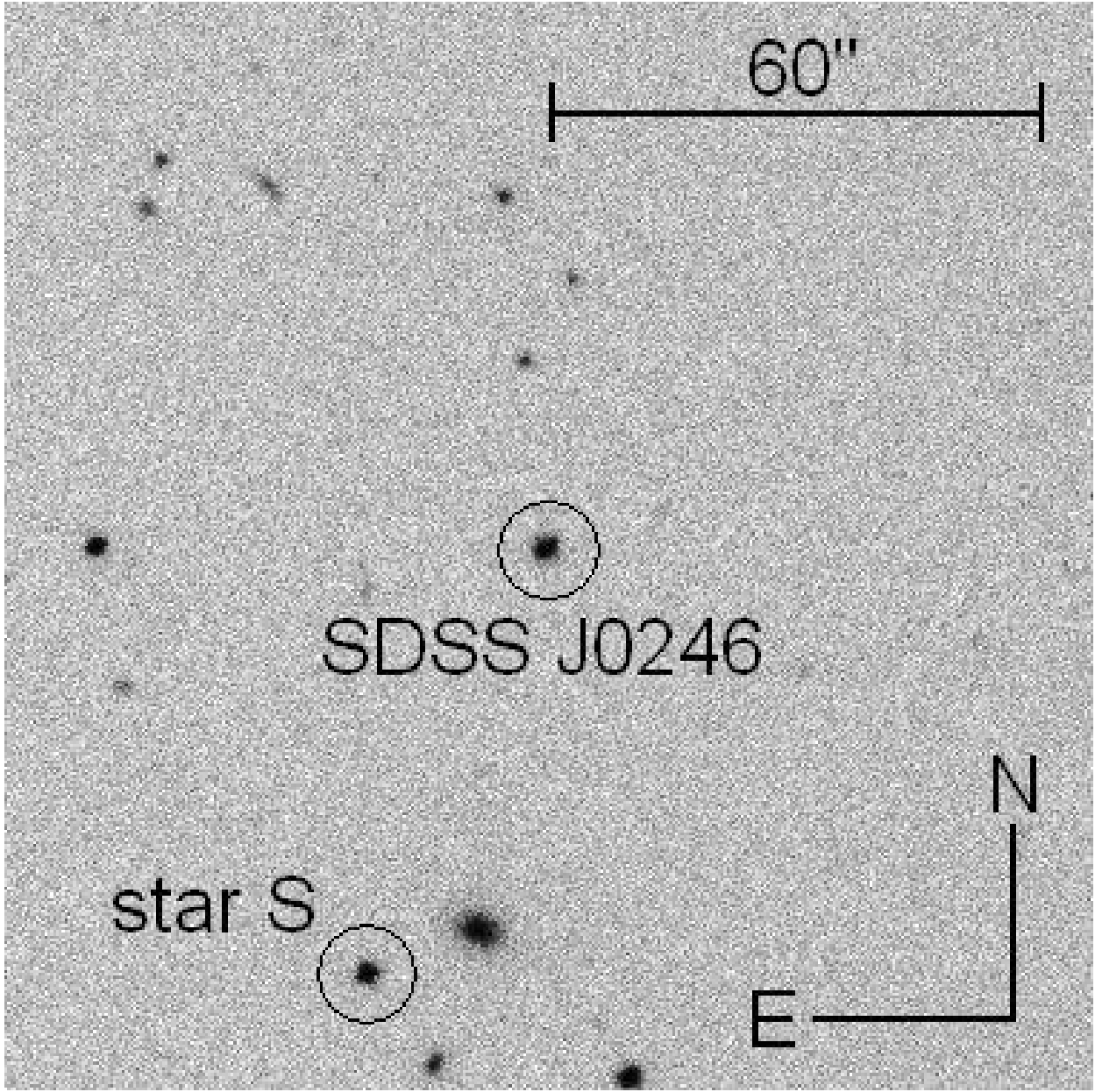}
\caption{SDSS $i$ band image of the field around SDSS J0246-0825.
Star S, which is detected in both the WB6.5m data
and the SDSS data, is used as a PSF template and a local 
photometric standard. 
\label{fig:field}}
\end{figure}

The $ugriz$ SDSS images of SDSS~J0246$-$0825 (sky and bias subtracted and
flat-field corrected) are shown in Figure~\ref{fig:sdss}. Two stellar
components are marginally-resolved/unresolved in the SDSS images; total
magnitudes (in $\sim$2\farcs0 aperture radius) of the two quasar components and the 
lensing galaxy (and an unknown extended object, see below) are 
18.21$\pm$0.02, 17.99$\pm$0.01,
17.95$\pm$0.01, 17.60$\pm$0.01, and 17.58$\pm$0.02 in $u$, $g$, $r$, 
$i$, and $z$, respectively. These errors in the magnitudes are 
statistical errors. We show the field image (SDSS $i$ band image) 
around SDSS~J0246$-$0825, including a nearby star (star S) 
which is seen in both the WB6.5m images and the SDSS images, in Figure
\ref{fig:field}. We use star S to estimate a Point Spread Function (PSF)
and to calibrate magnitudes. SDSS~J0246$-$0825 was targeted as a quasar 
candidate by the target selection pipelines, and the spectrum 
was taken by the SDSS spectrograph. The spectrum data are also unresolved 
due to the spatial resolution of the SDSS spectroscopy, but it clearly 
shows the quasar emission lines at $z=1.686\pm0.001$. 

\subsection{Follow-up Observation using the Walter Baade Telescope}

\begin{figure}
\plotone{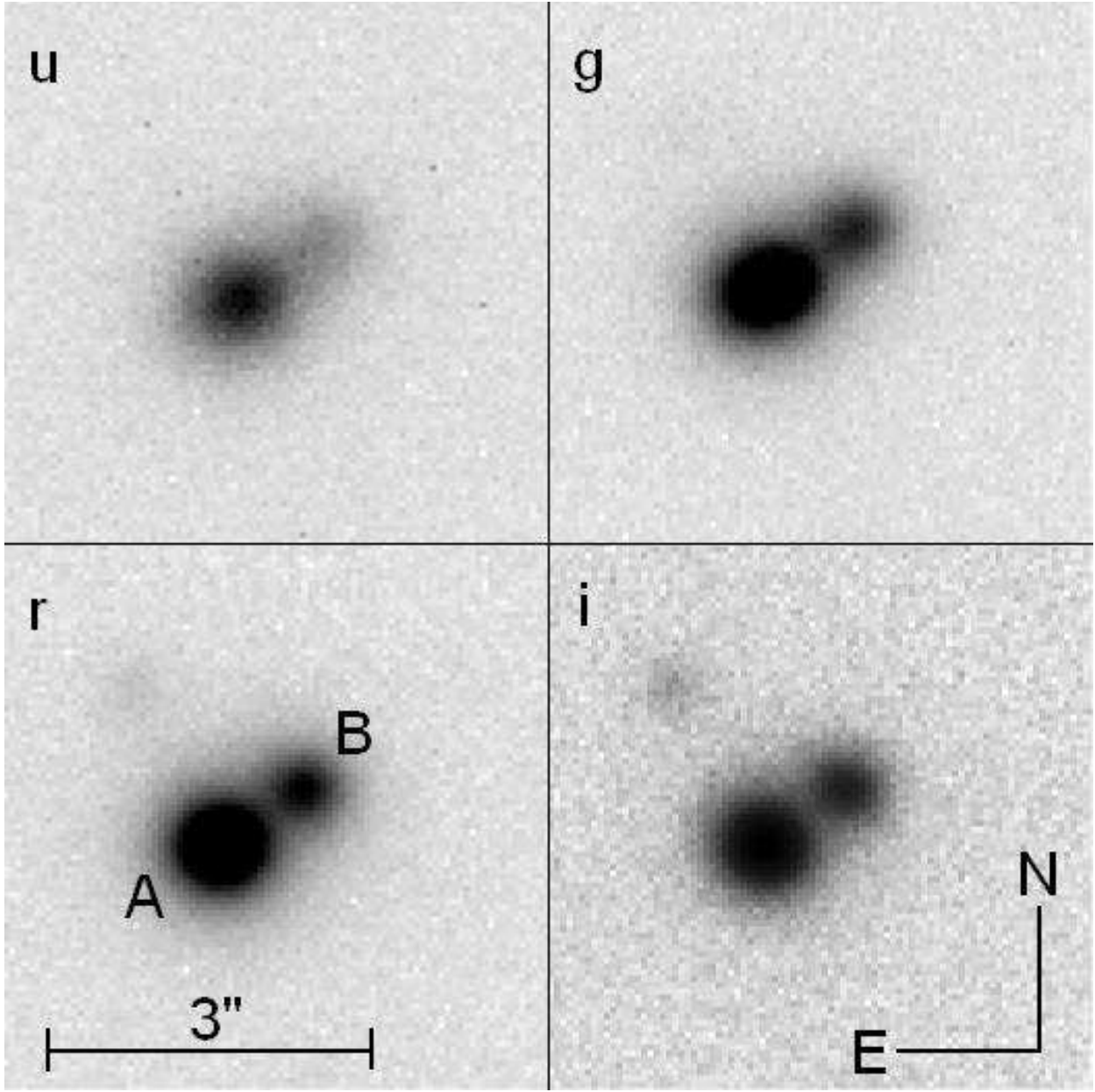}
\caption{The MagIC images of SDSS~J0246$-$0825 in $ugri$ filters
(appear as four separate images in this picture). Two stellar components 
are clearly seen in all images. The flux ratios between the two stellar 
components are 0.32, 0.31, 0.33, and 0.34 in $u$, $g$, $r$, and $i$,
respectively. The separation angle between A and B is 1\farcs042 in the
$i$ band image. The pixel size of these images is 0\farcs069, and the
seeing (FWHM) was less than 0\farcs6. The exposure time was 120 sec in each
image.  
\label{fig:mag_all}}
\end{figure}

High-resolution images of SDSS~J0246$-$0825 in the $u$, $g$, $r$, 
and $i$ filters were obtained with the Magellan Instant Camera 
(MagIC, a 2048$\times$2048 CCD camera; pixel scale is 0\farcs069)
on the WB6.5m telescope, on 2002 February 13.
The seeing (FWHM) was 
${\sim}$0\farcs5--0\farcs6, and the exposure time was 120 sec 
in each band. The $u$, $g$, $r$, and $i$ MagIC 
images of SDSS~J0246$-$0825 are shown in Figure \ref{fig:mag_all}. 
Two stellar components are clearly seen in all these images; 
we named the two stellar components A and B, with A
being the brighter component. The flux ratios between components B and
A (estimated by the single Gaussian fit) are 0.32, 0.31, 0.33, 
and 0.34 in $u$, $g$, $r$, and $i$, respectively. 

\begin{figure}
\plotone{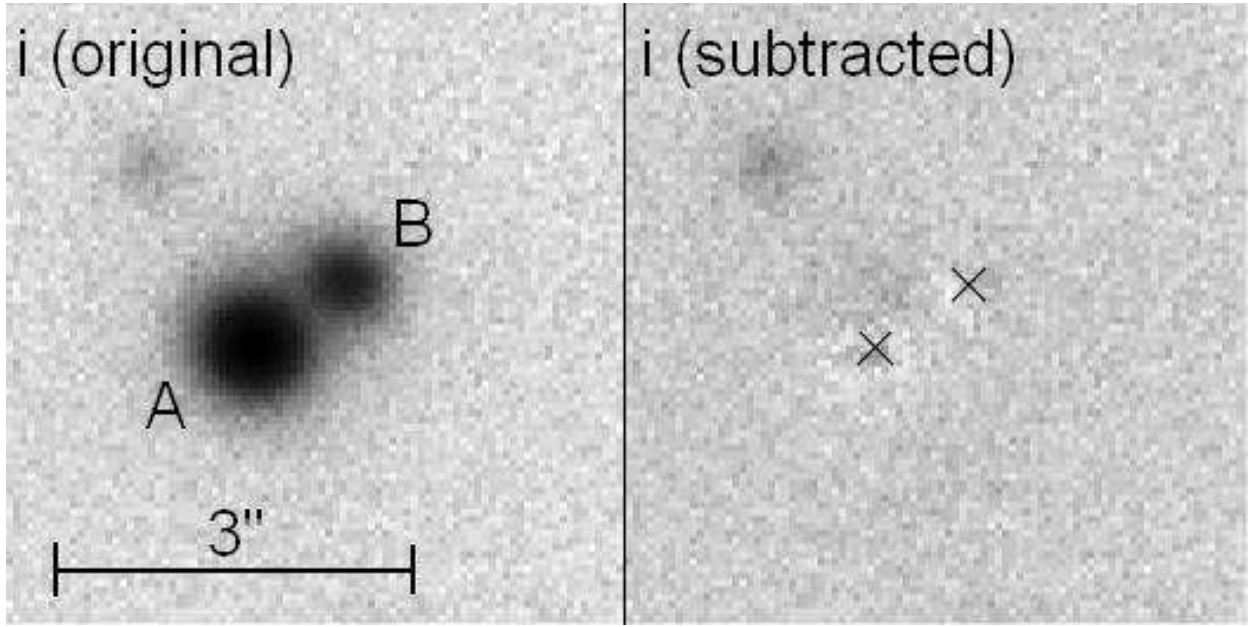}
\caption{The MagIC $i$ band image of SDSS~J0246$-$0825 ({\it left}) and
the PSF-subtracted image ({\it right}). The crosses in the
right panels represent measured centers of components A and B.
We find a faint, extended object (${\sim}$ 21.8 mag in 0\farcs5 aperture radius) 
between the two stellar components (on the line between the two stellar components), 
although it is not obvious in this figure. 
The PSF magnitudes of components A and B are calculated to be 17.7 and 18.9,
respectively.  
\label{fig:mag_i}}
\end{figure}

The PSFs derived from star S were subtracted from the
original $i$ band MagIC image.
The peak flux and the center coordinates 
of components A and B in the MagIC $i$ band image were calculated 
by the single Gaussian fit. The results agree with those obtained
by the imexamine task in IRAF\footnote[18]{IRAF is the Image Reduction
and Analysis Facility, a general purpose software system for the
reduction and analysis of astronomical data. IRAF is written and
supported by the IRAF programming group at the National Optical
Astronomy Observatories (NOAO) in Tucson, Arizona. NOAO is operated by
the Association of Universities for Research in Astronomy (AURA), Inc.
under cooperative agreement with the National Science Foundation.}. 
The $i$ band PSF-subtracted image is shown in the right panel of Figure
\ref{fig:mag_i}. Although it is not obvious in Figure \ref{fig:mag_i}, 
we find a faint, extended object (${\sim}$ 21.8 mag in 0\farcs5 aperture 
radius) between the two stellar components 
after the PSF subtraction. For further discussion of this object, 
see the Keck results in \S \ref{sec:keck} and/or 
the {\em HST} results in \S \ref{sec:hst}. 
The angular separation (in the $i$ band image) between components A and B is 
1\farcs042$\pm$0\farcs003. The $ugri$ band PSF magnitudes 
of components A and B and the $i$ band aperture magnitude of 
the extended objects between components A and B (component G, see below) 
are summarized in Table \ref{table:pos}. Here we used star 
S (see Figure \ref{fig:field}) as a photometric star\footnote[19]{
Note that there are small (a few percent) differences between the 
filter response functions of the WB6.5m and those of the SDSS.
}. 

\subsection{Follow-up Observation using the Keck Telescope}\label{sec:keck}

\begin{figure}
\plotone{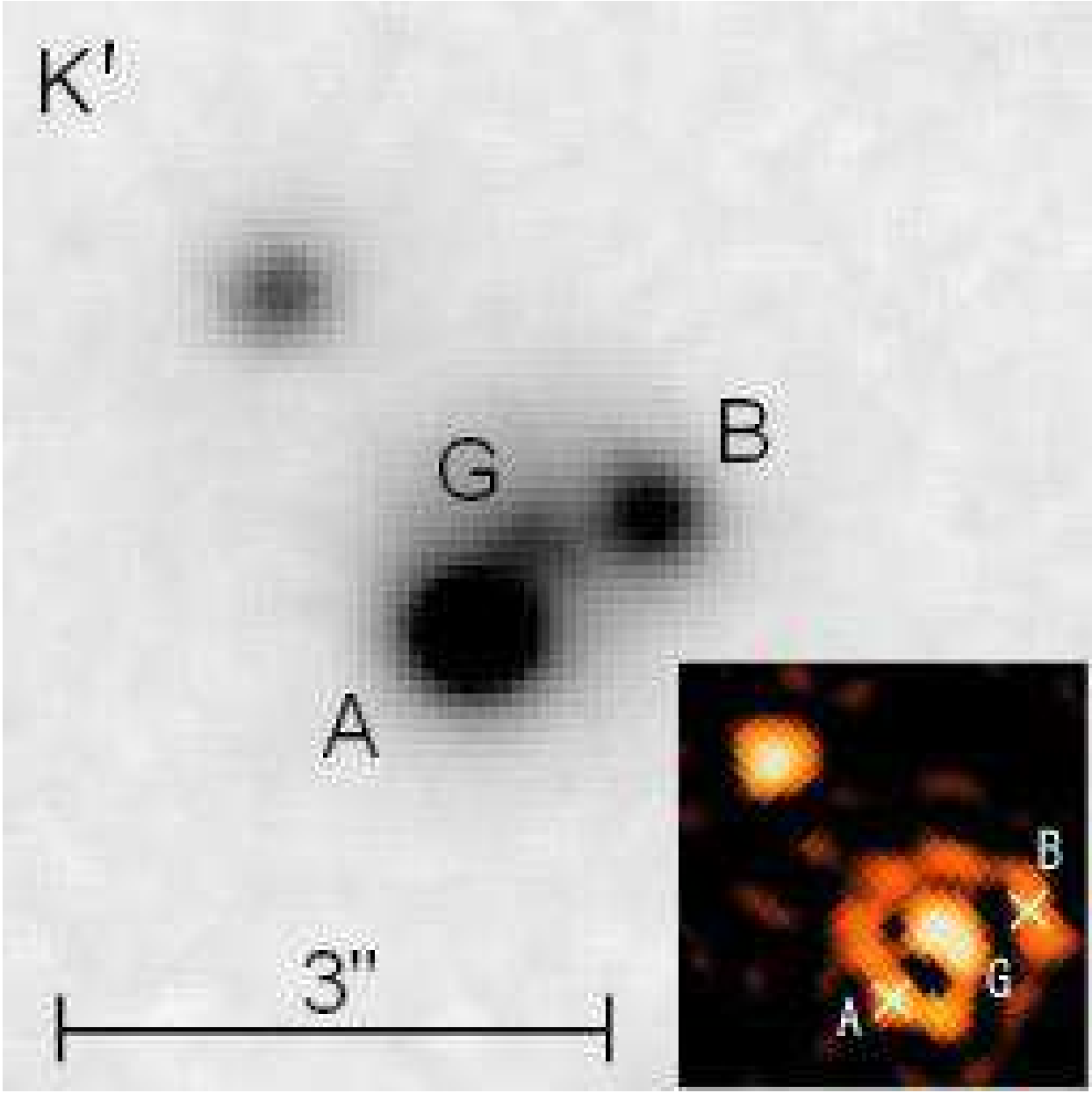}
\caption{The deconvolved NIRC $K'$ band image of SDSS~J0246$-$0825. 
We can clearly see an extended object (component G) between components A and B. 
The position of this object agrees with that of the extended object
found in the PSF-subtracted MagIC $i$ band image. The flux ratio between components 
B and A (0.29) is consistent with the average of the optical bands (0.33). In the inset, 
we show the PSF-subtracted image (the crosses represent measured centers of components A and B). 
\label{fig:keck-k}}
\end{figure}

We conducted a near-infrared imaging observation of SDSS~J0246$-$0825 
using the Keck I telescope, on 2002 August 21. Although the night 
was not photometric, we obtained $K'$ band data 
with the Near InfraRed Camera \citep[NIRC;][]{matthews94}. 
The seeing (FWHM) was variable, averaging 
about 0\farcs4, and the total exposure time was 900 sec. 
The deconvolved $K'$ band image is shown in 
Figure \ref{fig:keck-k}; the deconvolution was done using the
method which is described in \citet{magain98}. The pixel 
size is half that of the raw image, 0\farcs075. In 
Figure \ref{fig:keck-k}, we can clearly see an extended object (named component G)
at the same position of the extended object found in the 
PSF-subtracted MagIC $i$ band image (we can see component G more clearly in the 
PSF-subtracted images, the inset of Figure \ref{fig:keck-k}). 
This object is probably the lensing 
galaxy of this lensed quasar system; this hypothesis is further
supported by the spectroscopic observation of the two stellar components 
using the Keck telescope and the high-resolution imaging observation 
using the {\sl HST} (see \S \ref{sec:hst}).  
The flux ratio between components B and A is 0.29 in the $K'$ band image; 
this is consistent with the mean flux ratio of the optical bands. 

\begin{figure}
\plotone{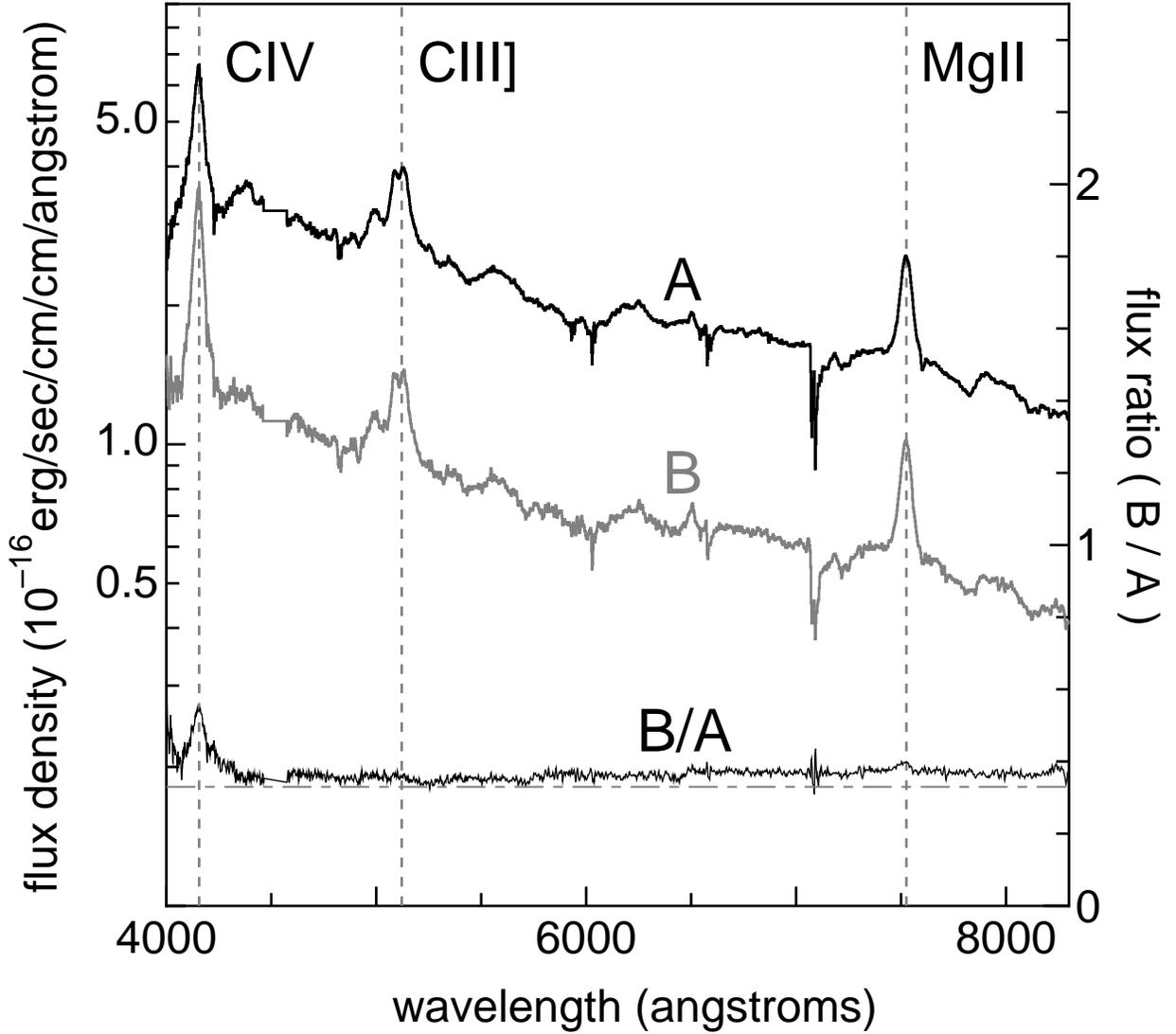}
\caption{The (binned) spectra of SDSS~J0246$-$0825 components A (black solid line) 
and B (gray solid line) taken with ESI on Keck II. The vertical grey dotted lines 
(4154.6 {\AA}, 5119.2 {\AA}, 7506.2 {\AA}) represent the positions of emission 
lines red-shifted to $z = 1.682$ of \ion{C}{4} (1549.06 {\AA}), 
\ion{C}{3]} (1908.73 {\AA}), and \ion{Mg}{2} (2798.75 {\AA}), respectively. 
We find both components have \ion{C}{4}, \ion{C}{3]}, and \ion{Mg}{2}
emission lines at the same redshift ($z=1.68$). 
The bottom thin black solid line represents the spectral flux ratio 
between components B and A. The grey horizontal dot-dashed line 
represents the mean flux ratio (0.33) found in $ugri$ photometry of the WB6.5m images.
The spectral flux ratio is almost constant and consistent with the photometric flux ratio. 
The data between 4460 {\AA} and 4560 {\AA} were excluded due to the existence of the 
bad columns. 
\label{fig:spec}}
\end{figure}

Spectra of the two components of SDSS~J0246$-$0825 were acquired 
using the Keck II telescope on
2002 December 4. We conducted a spectroscopic follow-up observation
using the Echellette Spectrograph and Imager \citep[ESI;][]{sutin97,sheinis02} with 
the MIT-LL 2048$\times$4096 CCD camera. We used the Echellette mode. 
The spectral range covers 3900{\,\AA} to 11,000{\,\AA}, and the spectral 
resolution is R $\simeq$ 27000. We set the slit width 1\farcs0 and the 
slit orientation so that the two 
stellar components (components A and B; see Figure \ref{fig:mag_all}) 
were on the slit (thus, component G should be also on the slit, but we were not 
able to extract the spectrum of component G because it is much fainter 
than the quasar components). The exposure time was 900 sec.
Since the seeing (FWHM) was about 0\farcs7 and the angular separation between the 
two components is only 1\farcs04, we can clearly see there are two objects on 
the slit but these are not well resolved. Therefore we extracted the spectra
using a method of summing the fluxes in a window around the position of the 
brighter component and in another window shifted by 1\farcs04. Sky was 
subtracted using neighboring windows on either side of the trace. 
There are bad columns on the ESI chip between 4460 {\AA} and 4560 {\AA}, and therefore,  
we excluded the data of this bad region from the spectra. 
The binned spectra of components A and B taken with ESI on 
Keck II are shown in 
Figure \ref{fig:spec}. In both spectra, \ion{C}{4}, \ion{C}{3]}, and
\ion{Mg}{2} emission lines are clearly seen at the same wavelength
positions. The redshifts of components A and B are 1.6820$\pm$0.0001 and 
1.6816$\pm$0.0004, which were calculated by fitting the Voigt functions 
with the \ion{C}{4} emission lines. The velocity difference between the
two quasar components is 20 km sec$^{-1}{\pm}40$ km sec$^{-1}$. 
The redshifts and the widths of the emission lines are summarized in
Table \ref{table:line}.  
In addition to the identical redshift, the spectral energy distributions
themselves are also quite similar. The bottom line in Figure \ref{fig:spec}
shows the spectral flux ratio between components B and A. 
We find that this is almost constant for a wide range of wavelengths and 
this is almost consistent with the mean flux ratio (0.33) of the $ugri$
images. A large difference is seen around the \ion{C}{4} emission 
line; this might be caused by microlensing of the broad emission line region 
\citep{richards04}.

\begin{figure}
\plotone{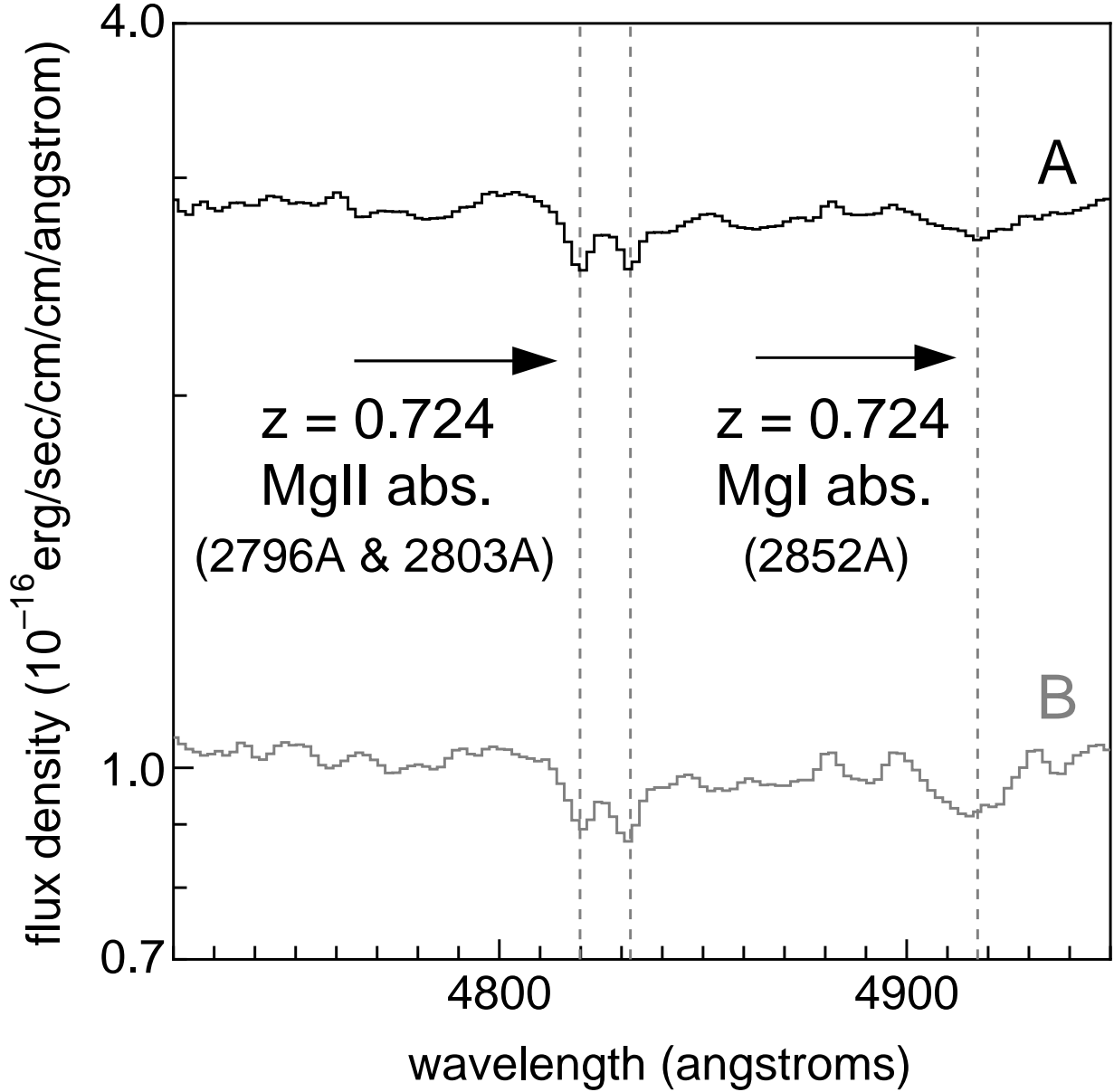}
\caption{An enlarged region of Figure \ref{fig:spec} showing the common
absorption line features in SDSS~J0246$-$0825 A and B. 
The vertical grey dotted lines (4819.5 {\AA}, 4831.8 {\AA}, and 4917.0
{\AA}) represent the positions of absorption lines red-shifted to
 $z=0.724$ of \ion{Mg}{2} (2795.53 {\AA} \& 2802.71 {\AA})  
and \ion{Mg}{1} (2852.13 {\AA}), respectively. The \ion{Mg}{2}
absorption lines and possibly the \ion{Mg}{1} absorption 
line are seen in both components. The rest frame equivalent widths are 
0.81 {\AA} and 0.62 {\AA} in the spectrum of component A, and 0.77 {\AA} and 0.79 {\AA}
in the spectrum of component B, for \ion{Mg}{2} 2795.53 {\AA} and  2802.71 {\AA}
absorption lines, respectively. 
\label{fig:abs}}
\end{figure}

In addition to the emission lines, we found a \ion{Mg}{2}/\ion{Mg}{1}
absorption system redshifted to $z=0.724$, as shown in Figure
\ref{fig:abs}. This absorption system probably arises from
a lensing galaxy, 
both because galaxies can produce such 
\ion{Mg}{2} absorption systems \citep{bergeron88,bergeron91} and because  
the \ion{Mg}{2}/\ion{Mg}{1} absorption system ($z=0.724$) is not so far
from the half of the angular diameter distance to the quasar components
($z=1.688$) that has the maximum lensing efficiency \citep{ofek03}. 
This prediction is further supported by the comparison of the expected 
magnitude by the mass models with the observed magnitude of the lensing 
galaxy (see \S \ref{sec:model}). Furthermore, we found two \ion{Mg}{2} 
absorption systems, whose redshifts are 1.537 and 1.531 
(both at $\sim$ 7100 {\AA}), with their \ion{Fe}{2} absorption lines 
(at $\sim$ 6600 {\AA}, $\sim$ 6000 {\AA}, and so on) and weaker absorption 
lines of other elements. Moreover, at $\sim$ 7100 {\AA}, 
we found two possible \ion{Mg}{2} absorption systems whose redshifts 
are 1.540 and 1.533, although their \ion{Fe}{2} absorption lines 
and other element absorption lines were not detected in the spectra. 
However these 4 absorption systems are 
unlikely to be associated with the lensing system, because the 
difference between the redshift of the source quasar and that of the 
absorption systems is small.

\subsection{Follow-up Observation using the {\sl HST}}\label{sec:hst}

Our final observation of SDSS~J0246$-$0825 was conducted using the
{\sl Hubble Space Telescope}, under the ``{\it HST} Imaging of
Gravitational Lenses'' program (principal investigator: C.~S.
Kochanek, proposal ID: Cycle12--9744). We used the Advanced Camera for 
Surveys \citep[ACS;][]{clampin00} and the Near Infrared Camera and 
Multi-Object Spectrometer \citep[NICMOS;][]{thompson92}, installed
on the {\sl HST}. The ACS and NICMOS
observations were conducted on 2003 October 13 and 2003 December 18,
respectively. We used F555W filter (${\approx}V$-band) and the F814W
filter (${\approx}I$-band) in the ACS observation and the F160W
filter (${\approx}H$-band) in the NICMOS observation. 

\begin{figure}
\plotone{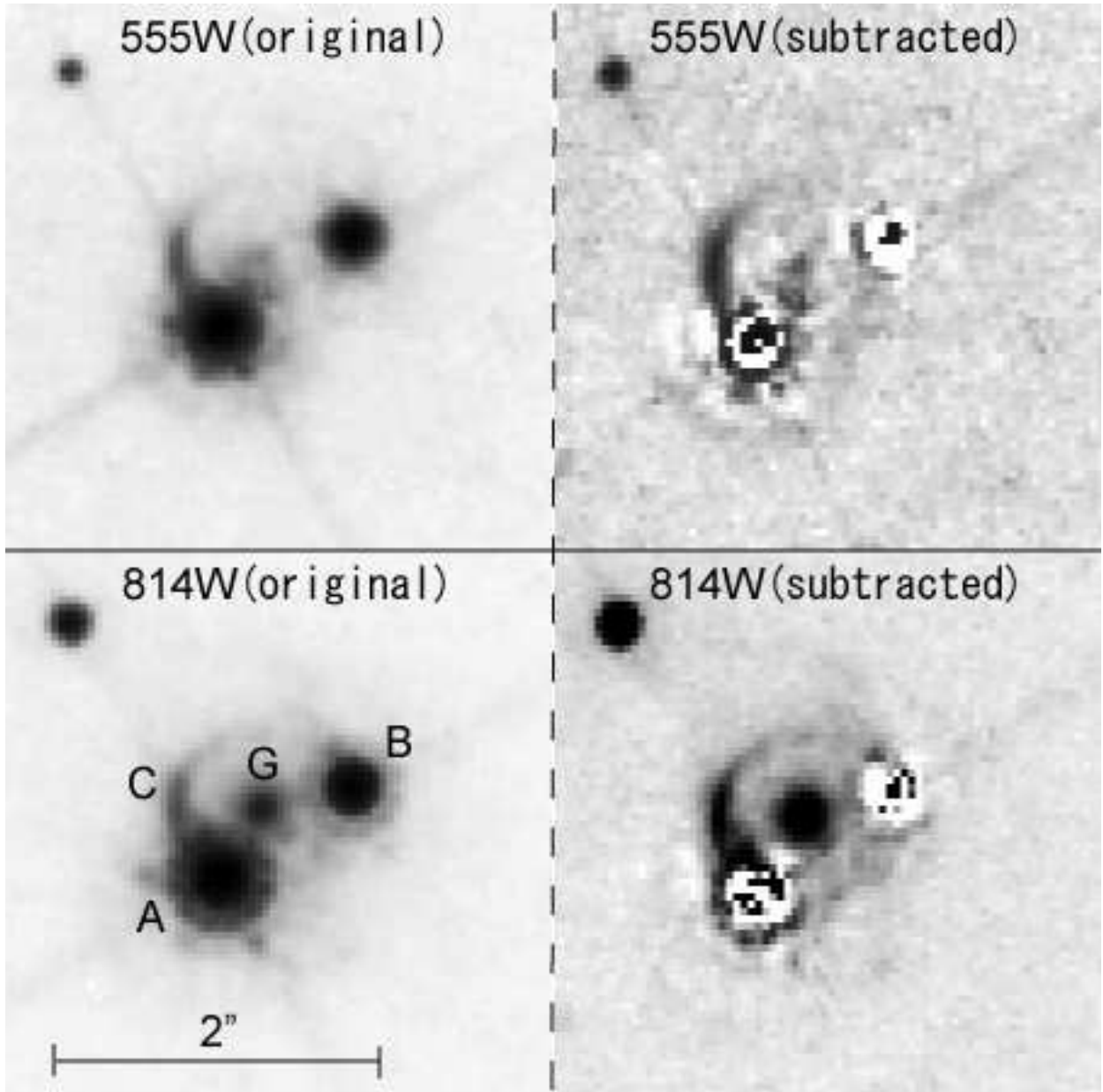}
\caption{
The ACS F555W image ({\sl upper left}) and F814W image ({\sl lower left}) of
SDSS~J0246$-$0825. Two quasar components are labeled A and B, and the
central extended object is labeled G. The pixel scale is approximately
0\farcs05 ${\rm pixel^{-1}}$. In addition to the central (lensing)
galaxy, we find a highly distorted object (component C) near component A. 
The PSF-subtracted images of F555W and F814W filters are shown in the right 
panels ({\sl upper right} and {\sl lower right}, respectively). Note that 
component A is saturated in both the F555W filter and F814W filter images. 
\label{fig:acs}}
\end{figure}

The ACS imaging observation consists of two dithered exposures taken in
ACCUM mode. Total exposure time of each dithered exposure for the F555W
filter was 1094 sec and that for the F814W filter was 1144 sec. 
The reduced (drizzled and calibrated) images were extracted using the
CALACS calibration pipeline which includes the
PyDrizzle algorithm. The combined image of the F555W filter and that of
the F814W filter are shown in the left panels of Figure~\ref{fig:acs}. 
As well as the NIRC $K'$ band image (Figure \ref{fig:keck-k}), we find the 
extended object (component G) between the two quasar components in the ACS 
images, particularly in the F814W image. The extended nature and the red 
color of component G suggest that this object is a galaxy. 
In addition to the identical nature of the spectral energy 
distributions of components A and B (described in \S \ref{sec:keck}),
the existence of the galaxy strongly supports the lensing
hypothesis; we conclude that SDSS~0246$-$0825 is certainly a lensed
quasar system. Although the spectrum of component G has not been obtained 
yet, the color of component G ($V-I{\approx}2.8$, estimated from the PSF-subtracted 
images, the right panels of Figure~\ref{fig:acs}) is consistent with it being 
a galaxy at $z=0.724$ \citep{fukugita95}, which is the redshift of 
\ion{Mg}{2}/\ion{Mg}{1} absorption system seen in the spectra of 
components A and B (see Figure~\ref{fig:abs}). 

In addition to the central extended object (component G), we find a
highly distorted object (named component C) in the upper left of
component A. Similar ``ring'' features of lensed (host) galaxies 
have been found in the optical and/or near-infrared band images 
\citep[e.g.,][]{impey98,king98,warren96},
and therefore this object is probably the lensed host galaxy of the
source quasar of SDSS~0246$-$0825. 
To improve the visibility of this distorted feature, 
we subtract components A and
B using the quasar PSFs produced by the Tiny Tim software \citep[version
6.1a;][]{krist03}. The PSFs of quasars were constructed with the Keck 
spectra of components A and B. The result is shown in
the right panels of Figure~\ref{fig:acs}. Component C is bright 
even in the F555W filter image as well as the F814W filter image 
(see Figure \ref{fig:acs}); the blue color of component C 
indicates that the host galaxy might be a starburst galaxy. The (magnification) 
center of component C is 0\farcs50 away from that of component A in the 
image plane, but the similar phenomenon have been observed in another 
lensed quasar system \citep{inada05}. See \S \ref{sec:model} for the 
comparison of this ``ring'' and mass models. 

\begin{figure}
\plotone{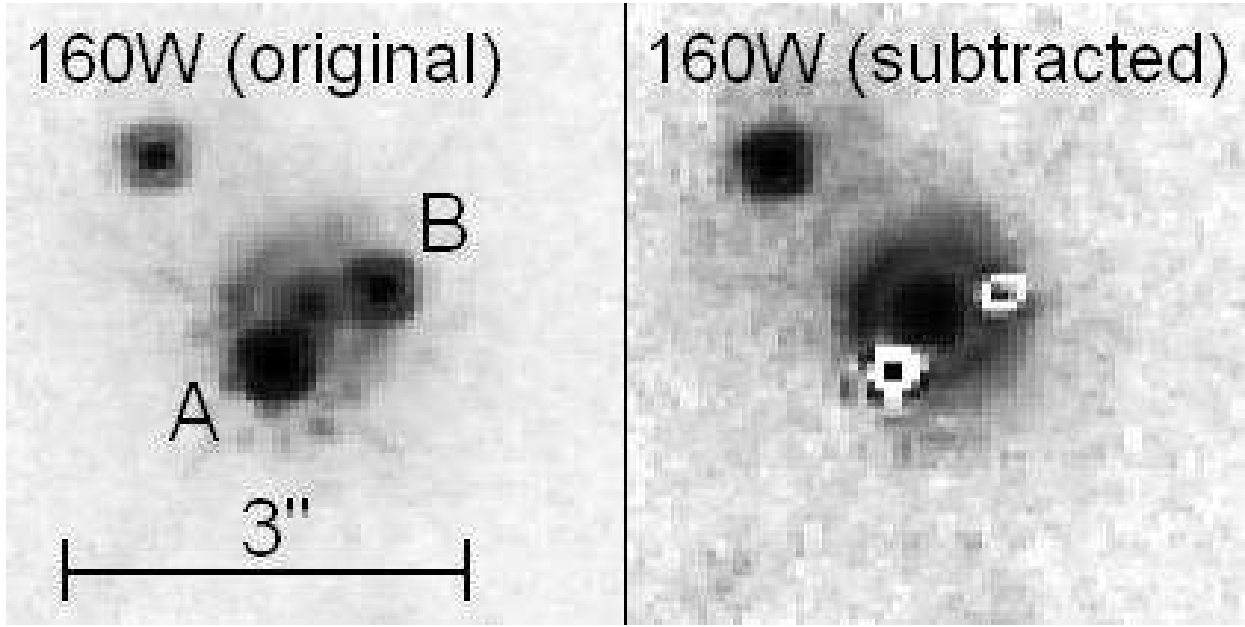}
\caption{
The NICMOS F160W image ({\sl left}) and the PSF subtracted image ({\sl
 right}) of SDSS~J0246$-$0825. The pixel scale is approximately
 0\farcs075 ${\rm pixel^{-1}}$. The central extended object (component G)
 is clearly seen in both the original and the PSF subtracted image. We can see 
 component C forms a nearly perfect ring in the PSF subtracted
 image. This ring feature is likely to be a lensed host galaxy of the 
 source quasar (see Figure \ref{fig:ring}). 
 The PSF magnitudes of components A and B are 17.3 and 18.6,
 respectively, and the aperture magnitudes (0\farcs375 radius) of
 component G is 19.4.  
\label{fig:nicmos}}
\end{figure}

The NICMOS imaging observation consists of four dithered exposures taken
in MULTIACCUM mode, through the F160W filter (${\approx}H$-band). 
The exposure time was 640 sec for three dithered exposures and 704 sec
for one dithered exposure. The reduced (calibrated) images were
extracted using the CALNICA pipeline. The central bad columns of each
dithered image were corrected by linear interpolation. 
The combined image is
shown in the left panel of Figure~\ref{fig:nicmos}. First, we confirm
the existence of the central extended object (component G) between components A and B,
and the highly distorted object (component C) near component A. 
In addition, we find a further highly distorted object, which is similar
to component C, near component B in the F160W image. To improve the visibility of them, 
we subtract components A and
B using the quasar PSFs produced by the Tiny Tim software. 
The PSF of a quasar was constructed with an
$\alpha_{\nu}=-0.5$ power law spectrum. The peak flux and the center
coordinates of components A and B in the NICMOS F160W image were
calculated using the imexamine task in IRAF. The result is shown in
the right panel of Figure \ref{fig:nicmos}; one can clearly see the
distorted objects form a nearly perfect ring. This prefect ring feature
can be also seen in the PSF-subtracted NIRC $K'$ band image 
(Figure \ref{fig:keck-k}). The PSF magnitudes of
components A and B and the aperture magnitude of component G in F160W
filter (${\approx}H$-band) are summarized in Table \ref{table:pos}.
In estimating the aperture magnitude of component G, we used 
the aperture radius of 0\farcs375 in the PSF subtracted image. 

\section{Mass Modeling}\label{sec:model}

\begin{figure}
\plottwo{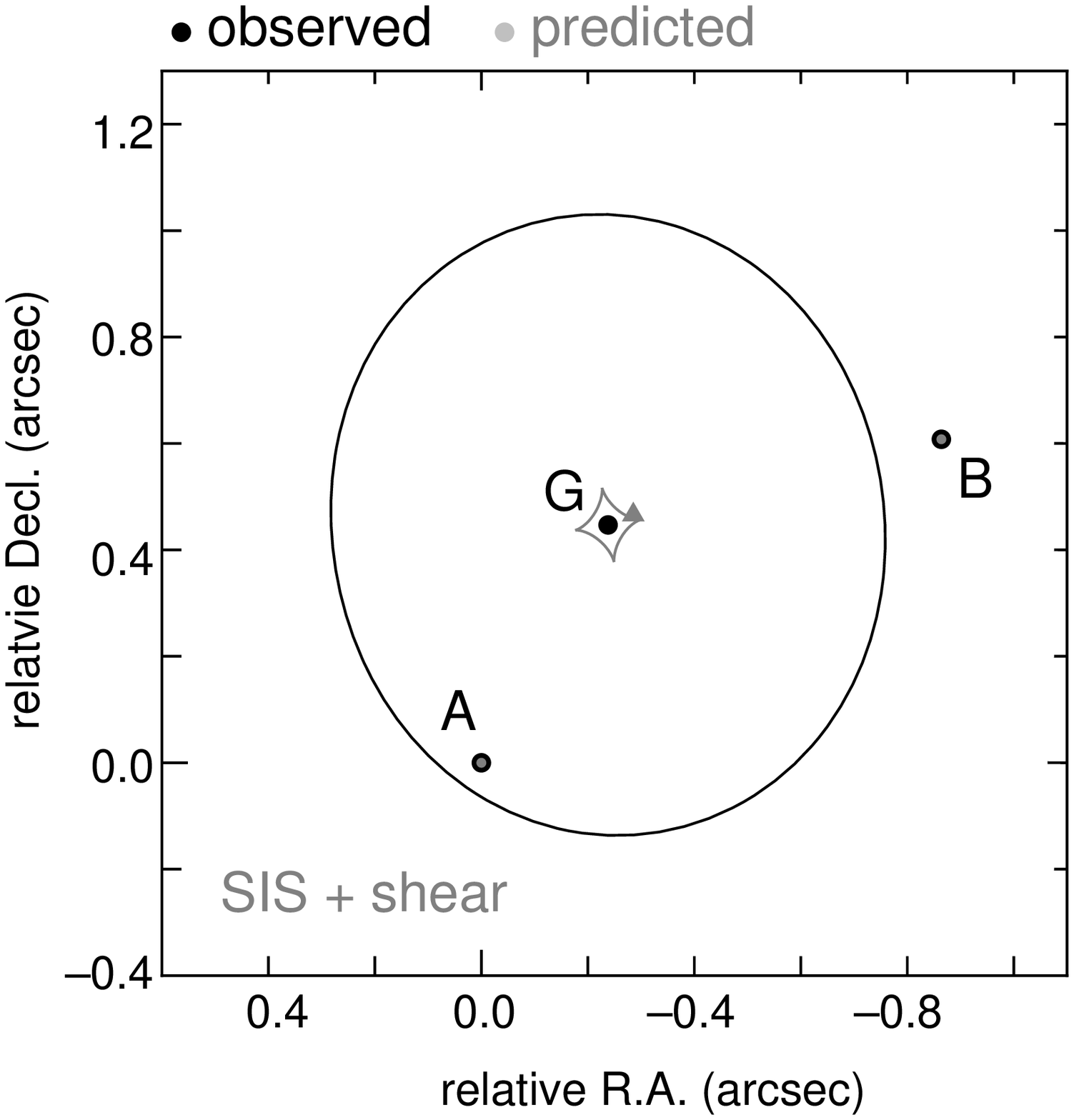}{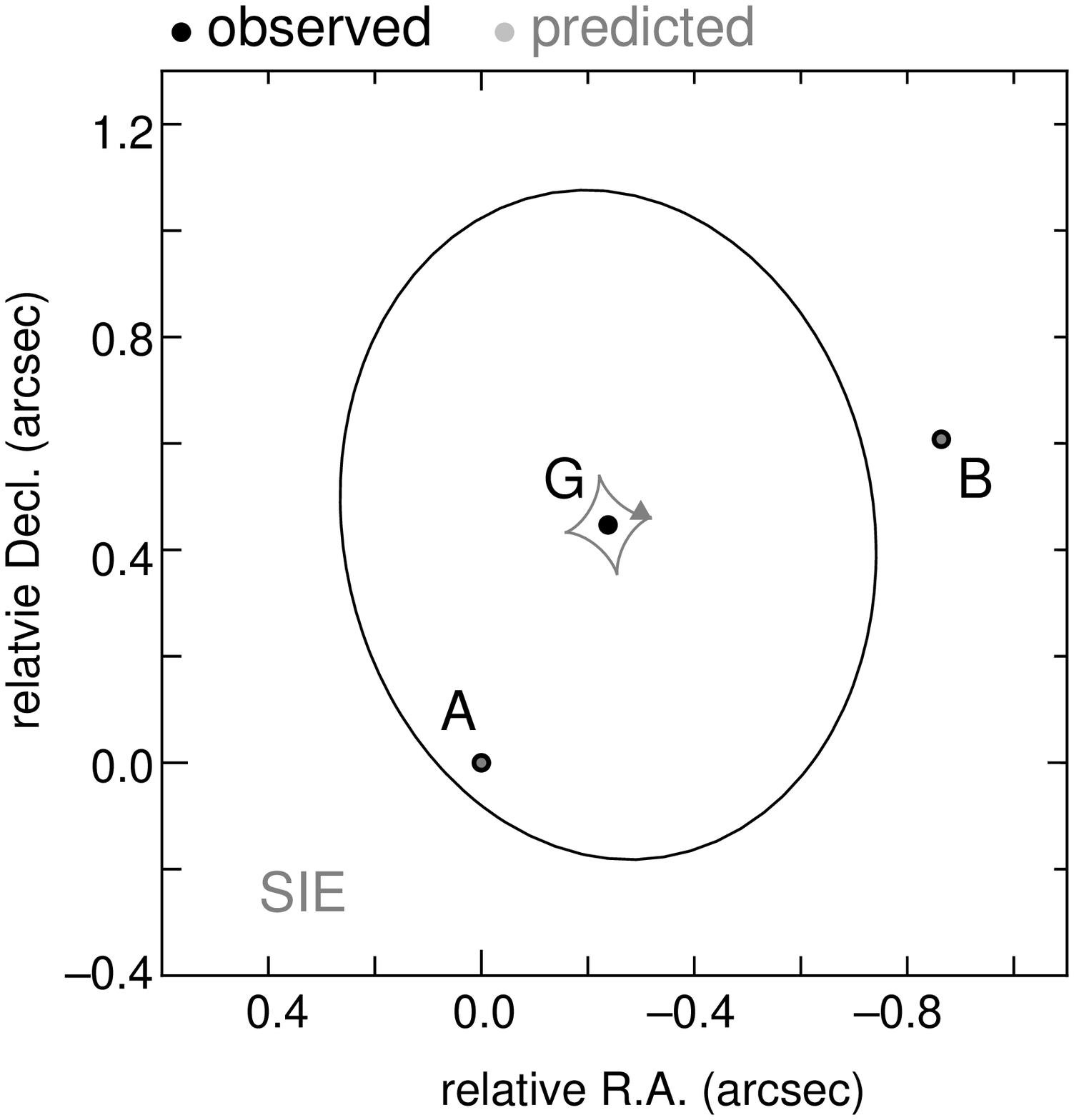}
\caption{
The critical curve (black solid lines) and the caustics (gray solid lines)  
of the best-fit SIS+shear model ({\em left} panel) and SIE model 
({\em right} panel). The black filled circles represent the
observed positions of the two lensed components (components A and B) and
the lensing galaxy (component G). The gray filled circles represent
the positions of the lensed components predicted by each mass model. 
The gray filled triangles are predicted source quasar positions. 
The best fit parameters for the SIS+shear model are 
${{\alpha}_{e}}=$0\farcs553,
$\gamma=0.067$, and $\theta_\gamma=10.147^\circ$ with the source quasar
position ($\Delta$R.A., $\Delta$Decl.) = ($-0\farcs285$, $0\farcs466$)
relative to component A. 
The best fit parameters for the SIE model are 
${{\alpha}_{e}}=$0\farcs554, $e=0.212$
($q=0.788$), and  ${\theta_e}=10.190^\circ$ with the source quasar
position ($\Delta$R.A., $\Delta$Decl.) = ($-0\farcs299$, $0\farcs469$)
relative to component A. Although the degree of freedom is zero in 
these models, both the mass models were not able to reproduce the observed 
flux ratio between components B and A.
\label{fig:model1}}
\end{figure}

We model the lensed quasar system using two mass models. One is the
Singular Isothermal Sphere with external shear (SIS+shear) model, with a
projected potential of 
\begin{equation}
\psi(r,{\theta}) = {{\alpha}_{e}}r + {\gamma \over 2} r^2 \cos2(\theta - \theta_\gamma), 
\end{equation}
where ${{\alpha}_{e}}$ is the Einstein radius in arcseconds, $r$ and
$\theta$ are the radial and the angular parts of the angular position on
the sky, respectively, and $\theta_\gamma$ is the position angle of the
shear, measured East of North. The other is the Singular Isothermal
Ellipsoid (SIE) model, with a projected potential of 
\begin{equation}
\psi(r,{\theta}) = {{\alpha}_{e}}r[1+((1-q^2)/(1+q^2))\cos 2(\theta-\theta_e)]^{1/2}, 
\end{equation}
where $q$ is the lens axis ratio and the $\theta_e$ is the position
angle of the ellipse, measured East of North. 
Since component A is saturated in both the ACS F555W image and the F814W 
image, we adopted the positions and flux
of components A and B in the MagIC image and the position of 
component G in the ACS (F814W) image. In order to obtain the
position of G relative to A, we used the positions of components B and 
G in the ACS F814W image and the position angle of the ACS image, 120.37 degree.
These observables are summarized in Table \ref{table:pos}. We use the
{\sl lensmodel} software \citep{keeton01b} to 
model the lensed quasar system. Both mass models have eight parameters;
${{\alpha}_{e}}$, $\gamma$, $\theta_\gamma$, the lensing galaxy
position, the source quasar position, and the source quasar flux for the 
SIS+shear model; ${{\alpha}_{e}}$, ellipticity $e$ ($e=1-q$), 
$\theta_e$, the lensing galaxy position, the source quasar position, and the source quasar flux 
for the SIE model. Since we
have only eight constraints (positions of two lensed components and the
lensing galaxy, and the fluxes of lensed components), the number of degrees of
freedom is zero. 
The results are summarized in Table \ref{table:model1} and Table \ref{table:model2}. 
The critical curve in
the image plane and the caustics in the source plane of the SIS+shear 
model are shown in the left panel of Figure \ref{fig:model1}, and those
of the SIE model are shown in the right panel of Figure \ref{fig:model1}, 
respectively. The best-fit models predict the time 
delay between the two lensed images to be ${\Delta}t=7.1h_{0}^{-1}$ day for
the SIS+shear model and ${\Delta}t=8.9h_{0}^{-1}$ day for the SIE model
($H_{0}=100h_{0}$ km sec$^{-1}$ Mpc$^{-1}$), assuming that the redshift of the
lensing galaxy is 0.724.  
Both the mass models predict the Einstein radius ${{\alpha}_{e}}$ of about
0\farcs55, which corresponds to the velocity dispersion (in one direction) 
of the lensing galaxy of $\sigma{\simeq}265$ km sec$^{-1}$ at $z=0.724$. 
Using the Faber-Jackson law \citep{faber76}, 
${M_{i}}^{\ast}=-21.3$ and ${\sigma}^{\ast}=225$ km
sec$^{-1}$ \citep{blanton01,kochanek96}, and  K- and evolution-correction
for elliptical galaxies (Inada 2004; 1.0 and $-$0.1, respectively, 
derived from the model of Poggianti 1997 and the SDSS filter response 
function), the $i$-band
magnitude of the lensing galaxy is predicted to be 22.1. This 
is in good agreement with the $i$-band magnitude of component
G, 21.8. Thus these mass models further support that the redshift of the 
lensing galaxy is 0.724. 

\begin{figure}
\plotone{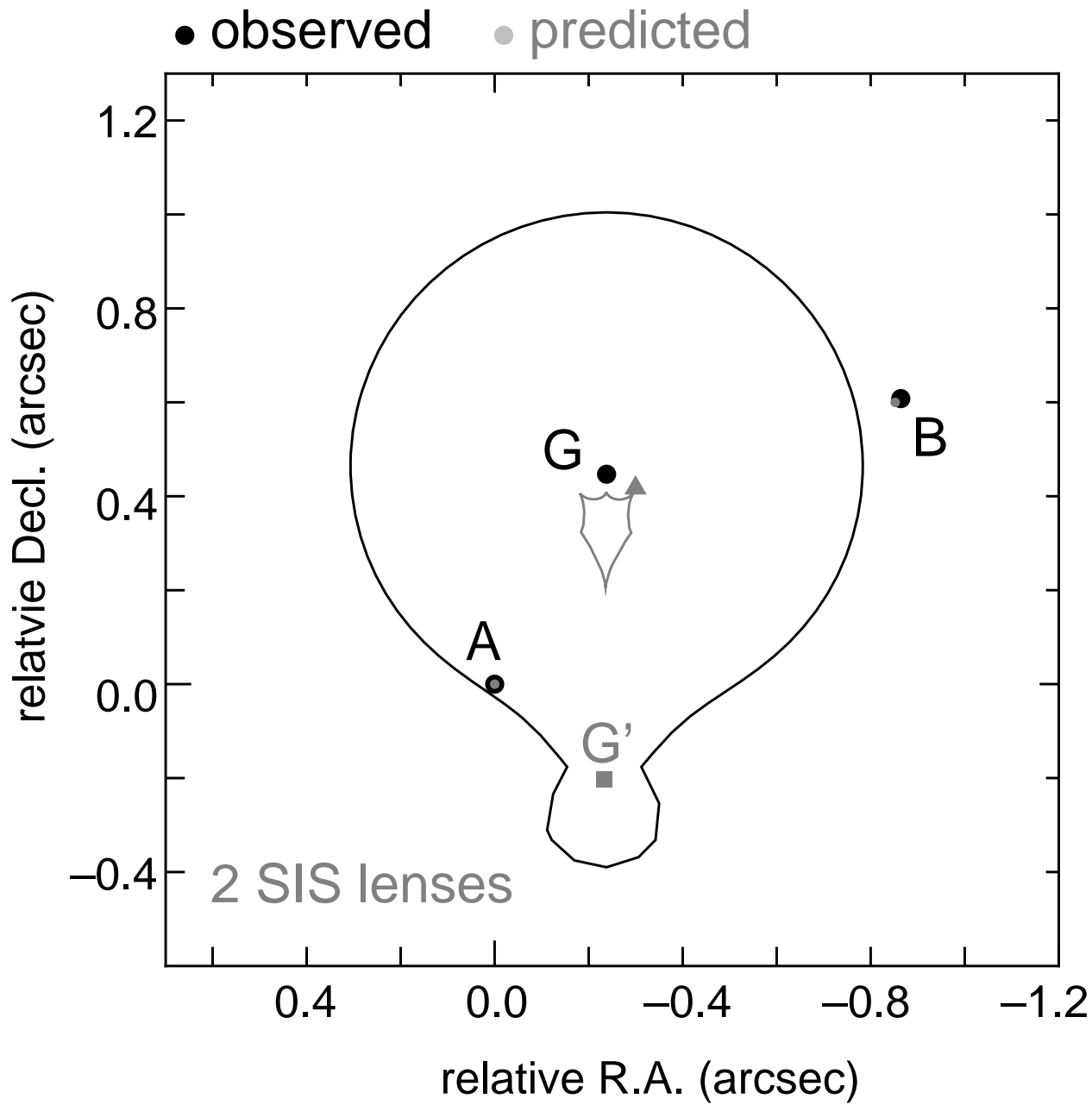}
\caption{
The critical curve (black solid lines) and the caustics (gray solid lines)  
of the 2 SIS lens model with ${\alpha}_{e}$(G$'$)$=$0\farcs07. 
The gray filled square represents 
the position of the predicted second lensing object G$'$. 
The best fit parameters of this model are 
${{\alpha}_{e}}=$0\farcs525 for component G, with the predicted 
position of G$'$ ($\Delta$R.A., $\Delta$Decl.)$=$($-0\farcs232$, $-0\farcs202$)
and the source quasar position 
($\Delta$R.A., $\Delta$Decl.)$=$($-0\farcs300$, $0\farcs418$), 
relative to component A. This model reproduces the observed flux ratio, 
B/A$=$0.33. 
\label{fig:model2}}
\end{figure}

Although the number of degrees of freedom is zero in the above mass models, 
we were not able to reproduce the observables perfectly. The observed
flux ratio between components B and A are $\sim 0.33$, while 
the predicted flux ratios are 0.58 for the SIS+shear model and 0.54 
for the SIE model (see Table \ref{table:model1}). 
The differences between the observed flux ratio and 
the predicted flux ratios are very large and therefore cannot be
explained by the observed data errors. To explore the discrepancy, 
we speculate that an unknown object near component G (we named this 
{\em hypothetical} second lensing object ``G$'$'') causes the discrepancy.
In order to minimize the difference between the numbers of constraints
and parameters, we simply applied the SIS model for both G and G$'$ 
(i.e., we did not consider the shear and the ellipticity). 
This model, however, has nine parameters and is under-constrained 
(the number of the difference between the constraints and the parameters 
is 1), thus we required that the Einstein radius of G$'$ should not be so 
large, smaller than $\sim$ 0\farcs20; this is natural since G$'$ should 
not be the dominant contribution. We varied $\alpha_e$ of G$'$
between 0\farcs01 and 0\farcs20, and search for the best-fit
models for each value of $\alpha_e$ of G$'$. 
We found that well-fitted models ($\chi^2\sim0$) exist for each $\alpha_e$ of G$'$ 
between $0\farcs05$ and $0\farcs20$. A result for the case that
the Einstein radius of G$'$ is 0\farcs07 is summarized in 
Table \ref{table:model1} and Table \ref{table:model2}; the predicted flux 
ratio between components B and A is 0.33. If we simply apply 
the Faber-Jackson law, 
the Einstein radius of the SIS model is proportional to $L^{1/2}$; 
thus that the Einstein radius of G$'$ is 0\farcs07 corresponds to 
G$'$ $\sim 5$ mag fainter than G if we assume that G$'$ is at the same 
redshift of G. This is consistent with the fact 
that we cannot see any objects at the predicted position of G$'$ in the
ACS images even after the PFS subtraction. 
The critical curve in the image plane and the caustics in the source plane of 
this model (2 SIS lens model for ${\alpha}_{e}$(G$'$)$=$0\farcs07)
are shown in Figure \ref{fig:model2}, and 
the predicted time delay of this model is ${\Delta}t=8.7h_{0}^{-1}$ day.   
We note that the total 
magnification factors of all mass models (especially 2 SIS lens model) 
are much higher than those of the previously known two-image lensed 
quasars, e.g., HE1104$-$1805 \citep{wisotzki93}. 

\begin{figure}
\plotone{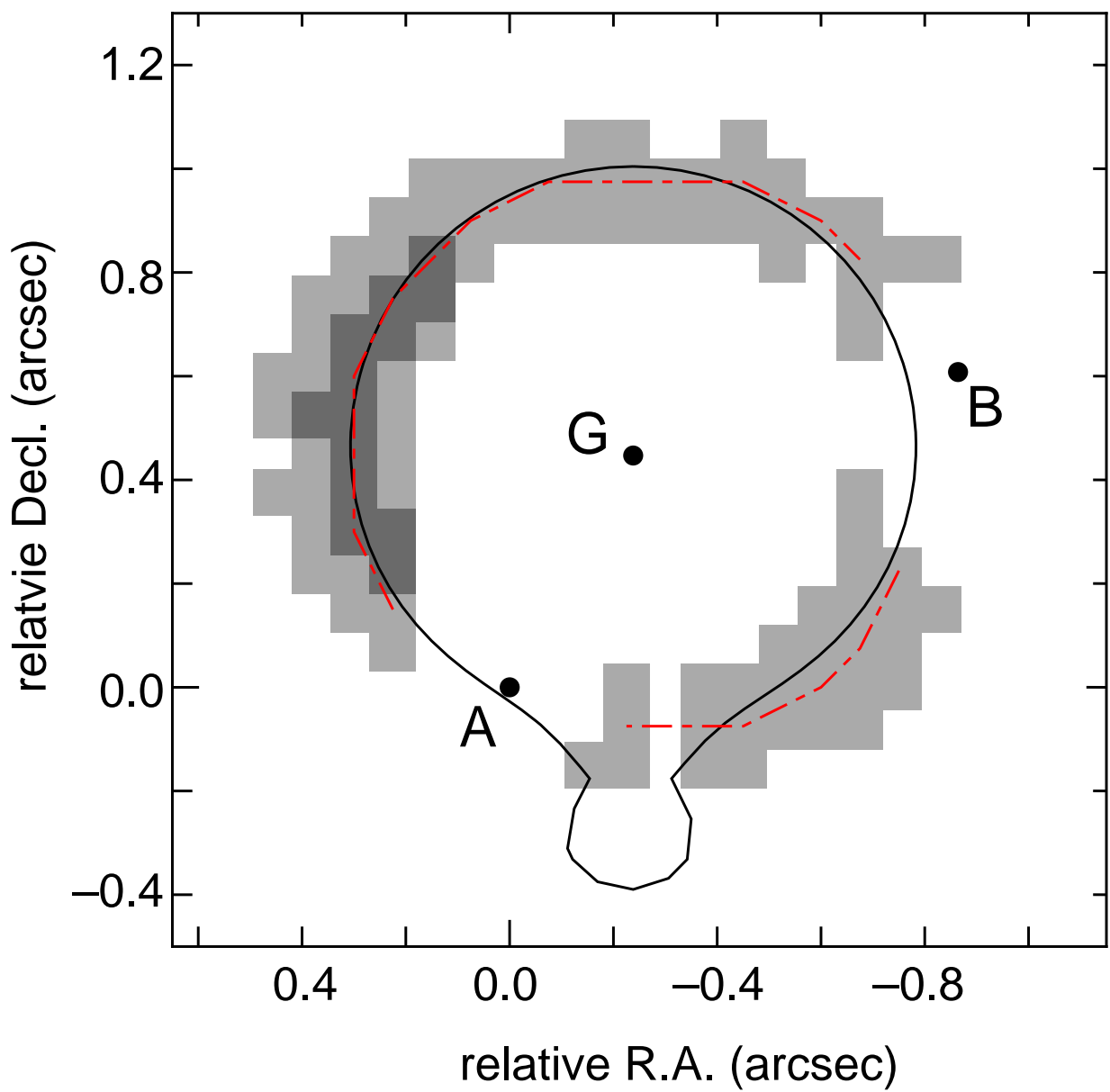}
\caption{
The comparison of the ``ring'' with the critical curve of the best-fit  
model. The solid line represents the critical curve
predicted by the 2 SIS lens model with ${\alpha}_{e}$(G$'$)$=$0\farcs07. The deep gray
squares represent the pixels with higher count rate, 
after subtracting the central lensing galaxy
and excluding the pixels around components A and B. The red
dot-dashed line connects the pixels which have the maximum count rate in
each column. The ring is in good agreement with the  
critical curve; this fact supports the idea that the ring is the lensed
host galaxy of the source quasar.  
\label{fig:ring}}
\end{figure}

In the above lens models, it is expected that the shape of the lensed
host galaxy, if it exists, is similar to that of the critical curve,
because the quadruple moments are negligible and the source is close to
the center of the lens system \citep{kochanek01}. To compare the 
ring feature observed in the NICMOS image and the critical curve
of the best-fit mass model, we overplot the critical curve predicted
by the 2 SIS lens model on the NICMOS ring, after subtracting
the central lensing galaxy and excluding the pixels around components A
and B. The result is shown in Figure \ref{fig:ring}. The red dot-dashed 
line in Figure \ref{fig:ring} connects the pixels that have the maximum 
count rate in each column; thus it almost corresponds to the structure
of the observed ring. In addition to the 2 SIS lens model, we also find 
that the critical curves of the SIS+shear model and the SIE model are in 
good agreement with the observed ring. 
A possible objection of this interpretation is that the
magnification center of the observed ring is largely different from the 
center of component A in both the ACS and the NICMOS images; 
the ring might not be associated to the source quasar but be a background 
galaxy lensed by the same lensing galaxy.

\section{Summary and Conclusion}\label{sec:conc}

We have reported the discovery of the doubly-imaged gravitationally
lensed quasar at $z=1.68$ with the separation angle 1\farcs04, 
SDSS~J0246$-$0825, which was selected from the SDSS data. 
The redshift of the lensing galaxy is likely to be
$z=0.724$, judged from the absorption lines in the spectra of the quasar
components and the apparent magnitude of the galaxy combined with the
expected absolute magnitude from the Faber-Jackson relation. 

We found that the simple mass models (SIS+shear and SIE) with
reasonable parameters well reproduce the lensing geometry of
SDSS~J0246$-$0825 but do not reproduce the flux ratio between the
lensed quasar components. Thus, we speculated that there is
another lensing object near the lensed quasar system; we found 
possible lens models which can reproduce both the 
lensing geometry and the flux ratio, by introducing a faint second 
lensing object. All of the best fit models (including the SIS + shear model 
and the SIE model) 
predicts the time delay between the two lensed images to be ${\sim}10h^{-1}$ days. 
In addition to the two lensed images, we have found a highly distorted
object in the {\sl HST} ACS and NICMOS images. In particular, in the
near infrared (NICMOS) image this object forms a nearly perfect ring. 
Since the critical curves of the best fit models show a good agreement with 
the ring, this is likely to be the lensed host galaxy of the source quasar. 

\acknowledgments 

A portion of this work was supported by NASA HST-GO-09744.20. N.~I. and
M.~O. are supported by JSPS through JSPS Research Fellowship for Young
Scientists. A portion of this work was also performed under the auspices 
of the U.S. Department of Energy,
National Nuclear Security Administration by the University of California,
Lawrence Livermore National Laboratory under contract No. W-7405-Eng-48. 
Some of the data presented herein were obtained at the W.M. Keck Observatory, 
which is operated as a scientific partnership among the California Institute 
f Technology, the University of California and the National Aeronautics and 
Space Administration. The Observatory was made possible by the generous 
financial support of the W.M. Keck Foundation. 

Funding for the creation and distribution of the SDSS Archive has been 
provided by the Alfred P. Sloan Foundation, the Participating Institutions, 
the National Aeronautics and Space Administration, the National Science Foundation, 
the U.S. Department of Energy, the Japanese Monbukagakusho, and the Max Planck 
Society. The SDSS Web site is http://www.sdss.org/. 

The SDSS is managed by the Astrophysical Research Consortium (ARC) for the 
Participating Institutions. The Participating Institutions are The University 
of Chicago, Fermilab, the Institute for Advanced Study, the Japan Participation 
Group, The Johns Hopkins University, the Korean Scientist Group, Los Alamos 
National Laboratory, the Max-Planck-Institute for Astronomy (MPIA), the 
Max-Planck-Institute for Astrophysics (MPA), New Mexico State University, 
University of Pittsburgh, University of Portsmouth, Princeton University, 
the United States Naval Observatory, and the University of Washington.

\clearpage

\clearpage

\clearpage

\clearpage

\clearpage

\clearpage

\clearpage

\clearpage

\clearpage

\clearpage

\clearpage

\clearpage

\clearpage

\begin{deluxetable}{crrrrrrr} 
\rotate
\tablecolumns{9} 
\tablewidth{0pc} 
\tablecaption{POSITIONS AND MAGNITUDES OF COMPONENTS OF SDSS J0246$-$0825} 
\tablehead{ 
\colhead{Object} & \colhead{${\Delta}$R.A.($^{''}$) \tablenotemark{a}} & 
\colhead{${\Delta}$Decl.($^{''}$) \tablenotemark{a}} & 
\colhead{$u$ \tablenotemark{b}} & 
\colhead{$g$ \tablenotemark{b}} &
\colhead{$r$ \tablenotemark{b}} &
\colhead{$i$ \tablenotemark{b}} &
\colhead{F160W} }
\startdata 
Component A &  0.000{$\pm$}0.001   & 0.000{$\pm$}0.001 & 19.17{$\pm$}0.05 & 18.73{$\pm$}0.02 & 18.34{$\pm$}0.02 & 17.94{$\pm$}0.03 & 17.3{$\pm$}0.1 \\
Component B & $-$0.852{$\pm$}0.001 & 0.600{$\pm$}0.001 & 20.41{$\pm$}0.10 & 20.00{$\pm$}0.03 & 19.55{$\pm$}0.03 & 19.11{$\pm$}0.04 & 18.6{$\pm$}0.2 \\
Component G & $-$0.238{$\pm$}0.002 & 0.447{$\pm$}0.002 & \nodata\phn      & \nodata\phn      & \nodata\phn      & 21.80{$\pm$}0.60 & 19.3{$\pm$}0.5 \\
\enddata
\tablenotetext{a}{The celestial coordinates of component A is 024634.11$-$082536.3, J2000.}
\tablenotetext{b}{The errors do not include the photometric uncertainty of the standard star.}
\label{table:pos}
\end{deluxetable}

\clearpage

\begin{deluxetable}{lccccccc} 
\tablecolumns{8} 
\tablewidth{0pc} 
\tablecaption{EMISSION LINES} 
\tablehead{ 
\colhead{} & \multicolumn{3}{c}{Component A} & \colhead{} & 
\multicolumn{3}{c}{Component B} \\ 
\cline{2-4} \cline{6-8} \\ 
\colhead{Element({\,\AA})} & \colhead{${\lambda}_{obs}$({\,\AA})} & \colhead{FWHM({\,\AA})} & \colhead{Redshift} & \colhead{} &
\colhead{${\lambda}_{obs}$({\,\AA})} & \colhead{FWHM({\,\AA})} & \colhead{Redshift} }
\startdata 
\ion{C}{4}(1549.06)  & 4154.56 & 69.5 & 1.6820$\pm$0.001 & & 4153.97  & 68.7 & 1.6816$\pm$0.004 \\
\ion{C}{3]}(1908.73) & 5116.82 & 85.1 & 1.6807$\pm$0.003 & & 5117.09  & 87.6 & 1.6809$\pm$0.004 \\
\ion{Mg}{2}(2798.75) & 7524.94 & 65.8 & 1.6887$\pm$0.002 & & 7524.07  & 64.7 & 1.6884$\pm$0.004 \\
\enddata 
\label{table:line}
\end{deluxetable} 

\clearpage

\begin{deluxetable}{ccccccccc}
\tablecolumns{9} 
\tablewidth{0pc} 
\tablecaption{RESULT OF LENS MODELING: I.} 
\tablehead{ 
\colhead{Model} & 
\colhead{$\alpha_e$} & 
\colhead{$\gamma$ or $e$} & 
\colhead{$\theta_{\gamma}$ or $\theta_{e}$} & 
\colhead{${\Delta}$R.A.(source)} & 
\colhead{${\Delta}$Decl.(source)} & 
\colhead{{$\mu_{\rm tot}$} \tablenotemark{a}} &
\colhead{B/A \tablenotemark{b}} &
\colhead{$\chi^{2}$ \tablenotemark{c}} }
\startdata 
SIS+shear    &  0\farcs553  &  0.067  &  10.147$^\circ$  &  $-$0\farcs285  &  0\farcs466 & 13.9 & 0.58 & 4.70 \\
SIE          &  0\farcs554  &  0.212  &  10.190$^\circ$  &  $-$0\farcs299  &  0\farcs469 & 12.4 & 0.54 & 3.64 \\
2 SIS lenses &  0\farcs525  &  \nodata\phn  &  \nodata\phn  &  $-$0\farcs300  &  0\farcs418 & 32.7 & 0.33 & 0.16 \\
\enddata
\tablenotetext{a}{Predicted total magnification.}
\tablenotetext{b}{Flux ratio between B and A.}
\tablenotetext{c}{$\chi^{2}$ of model fitting. In model fitting, we assume slightly larger errors for the flux ratio (B/A) 
than the observed errors; 10\% errors for SIS+shear and SIE and 5\% errors for 2 SIS lenses. 
}
\label{table:model1}
\end{deluxetable}

\begin{deluxetable}{ccccccc}
\tablecolumns{7} 
\tablewidth{0pc} 
\tablecaption{RESULT OF LENS MODELING: II.} 
\tablehead{ 
\colhead{Model} & 
\colhead{${\Delta}$R.A. (G)} & 
\colhead{${\Delta}$Decl.(G)} & 
\colhead{$\alpha_e$ (G$'$)} & 
\colhead{${\Delta}$R.A. (G$'$)} & 
\colhead{${\Delta}$Decl.(G$'$)} }
\startdata 
SIS+shear    &  0\farcs238  &  0\farcs447  &  \nodata\phn    & \nodata\phn & \nodata\phn  \\
SIE          &  0\farcs238  &  0\farcs447  &  \nodata\phn     & \nodata\phn & \nodata\phn  \\
2 SIS lenses &  0\farcs238  &  0\farcs447  &  0\farcs070   &  $-$0\farcs232  &  $-$0\farcs202 \\
\enddata
\label{table:model2}
\end{deluxetable}

\end{document}